\documentclass[runningheads]{llncs}
\usepackage{graphicx}
\usepackage{multirow}
\usepackage{booktabs} 
\usepackage{mathtools}
\usepackage{subcaption}
\usepackage{amsfonts}
\usepackage{tikz}
\usepackage{pdflscape}
\usepackage{verbatim}
\usepackage{algorithm}
\usepackage{algorithmic}
\usepackage{afterpage}

\newcommand{\quotes}[1]{``#1''}

\usetikzlibrary{arrows, positioning, shadows, calc,trees,backgrounds,shapes,decorations.pathmorphing,fit,positioning,chains}
\def\lav{gray!60}
\def\oran{gray!20}
\tikzstyle{peers}=[draw,circle,fill=\lav, \lav, minimum width=15pt, thin, align=center, anchor=base, font=\fontfamily{pzc}\selectfont, text=black]
\tikzstyle{superpeers}=[draw, minimum height=1.6em, minimum width=1em, fill=\oran, double copy shadow={shadow xshift=1pt, shadow yshift=1pt, fill=white, draw}, draw, rectangle, \oran, thin, align=center, anchor=base, font=\fontfamily{pzc}\selectfont, text=black]
\tikzstyle{legendsp}=[rectangle, draw, rounded corners, thin,fill=\oran, \oran, minimum width=2cm,align=center, anchor=base, font=\fontfamily{pzc}\selectfont, text=black]
\tikzstyle{legendp}=[rectangle, draw, rounded corners, thin, fill=\lav, \lav, minimum width= 2cm,align=center, anchor=base, font=\fontfamily{pzc}\selectfont, text=black]

\begin{document}

\title{Impact of the Query Set on the Evaluation of Expert Finding Systems}
%
%
\author{Robin Brochier\inst{1,2} \and
Adrien Guille\inst{1} \and
Benjamin Rothan\inst{2} \and
Julien Velcin\inst{1}}
\authorrunning{R. Brochier et al.}
%
\institute{Universit\'e de Lyon, Lyon 2, ERIC EA 3083, France \\
\email{\{robin.brochier,julien.velcin, adrien.guille\}@univ-lyon2.fr}\\
\url{https://eric.ish-lyon.cnrs.fr/}\and
Digital Scientific Research Technology, Lyon, France
\email{\{robin, benjamin\}@peer.us}\\
\url{https://peer.us/}}
\maketitle              
\begin{abstract}
Expertise is a loosely defined concept that is hard to formalize. Much research has focused on designing efficient algorithms for expert finding in large databases in various application domains. The evaluation of such recommender systems lies most of the time on human-annotated sets of experts associated with topics. The protocol of evaluation consists in using the namings or short descriptions of these topics as raw queries in order to rank the available set of candidates. Several measures taken from the field of information retrieval are then applied to rate the rankings of candidates against the ground truth set of experts. In this paper, we apply this \textit{topic-query} evaluation methodology with the AMiner data and explore a new \textit{document-query} methodology to evaluate experts retrieval from a set of queries sampled directly from the experts documents. Specifically, we describe two datasets extracted from AMiner, three baseline algorithms from the literature based on several document representations and provide experiment results to show that using a wide range of more realistic queries provides different evaluation results to the usual \textit{topic-queries}.      

\end{abstract}
\section{Introduction}

It is common to consider expertise as an implicit knowledge about a domain that someone carries and shares in different manners. Expertise retrieval aims at identifying this knowledge through explicit artifacts such as communications, actions or interactions between people. When someone call for an expert, she expects to find a candidate able to understand a specific query.
Whereas most evaluations for expertise retrieval consist in directly querying the namings or descriptions of the ground truth topics of a given dataset, we claim that these queries do not show much interest for a real case scenario since:
\begin{itemize}
\item the textual content of the topics namings are very limited in terms of language. Using richer (hence noisier) descriptions might better test the robustness of the evaluated algorithms. For example, it is better to query multiple times a retrieval algorithm with several texts relevant to the field of \quotes{data mining} than only once with the naming of the field itself. In real case scenarios, users have a wide range of behaviors and seldom use the same queries when looking for the same thing     
\item no one really seeks for experts in so broad subjects. Most of the time, someone looks for an expert with a very specific application in mind. Indeed, if a recruiter from a company is looking for a researcher to work on a specific subject, it is more likely that she will use the detailed description of the project instead of a generic naming of the job to find the right person. 
\end{itemize}      

In this paper, we first provide in Section \ref{sec:definition} a formal definition of the expert finding task  applied to the data extracted from AMiner \footnote{http://AMiner.org/}. In particular, we describe two protocols: the \textit{topic-query} evaluation and the \textit{document-query} evaluation. We then describe in Section \ref{sec:baseline} three baseline algorithms from the literature that we reimplemented and tested using several document representations. Finally we show and analyze in Section \ref{sec:experiments} the results of our experiments, demonstrating the impact of the type of query on the behaviors of the algorithms and document representations.      

Precisely, our contribution is fourfold:
\begin{enumerate}
\item we propose two different procedures for generating queries and study their impact on the evaluation results  
\item we describe two ways of using AMiner's data for expert finding and detail the preprocessing needed  
\item we reimplement and evaluate 3 algorithms from the literature based on several document representations
\item the corresponding Python code is made publicly available \footnote{https://github.com/brochier/impact\_query\_expert\_finding} which makes it easy to reproduce the experiments or even expand the proposed pipeline.    
\end{enumerate}

\section{Related Works}

The automation of expert finding appeared as a research field along with the creation of large databases when started the digitalization of libraries and of the communication tools in big companies. \textit{P@noptic Expert} \cite{craswell2001p} is one of the first published works on expertise retrieval. The proposed model transforms the expert finding task in a text similarity task by building a meta-documents for each candidate, aggregating all documents where the name of this candidate appears. In 2005, the research around expert finding received a boost with the \textit{TREC-2005 Enterprise Track, Expert search task}. They provided a dataset extracted from the \textit{World Wide Web Consortium (W3C)}. Moreover, they shared an evaluation toolkit to allow researchers to confront their algorithms. As a result, a formal definition of the problem emerged \cite{craswell2005overview}. As presented in \cite{balog2006formal}, the generative document-model of Balog et al., we denote $q$ a query, $d$ a document and $e$ a candidate. The expert finding task consists in estimating the probability of a candidate to be an expert given a query $P(e|q) = \frac{P(q|e)P(e)}{P(q)}$. Voting models as in \cite{macdonald2006voting} relax the probabilistic view of the latter equation. As an example, the score of a candidate can be computed by ranking all documents against the query with a document representation such as the bag-of-words based model term frequency. Then each candidate is provided a score given the ranks of the documents she is associated to. In \cite{zhang2007expert} and \cite{serdyukov2008modeling}, the authors propose to propagate the affinity between the query and the documents across the collaboration graph in a similar manner as PageRank \cite{page1999pagerank}. More recently, \cite{van2016unsupervised} adapted a word embedding technique to embed words and candidates in the same vector space. 
Many algortihms presented recently in the field of representation learning such as TADW \cite{yang2015network} and metapath2vec \cite{dong2017metapath2vec} can be adapted to the task of expert finding but their authors did not experiment them on this specific task. Much work has been done for expert finding in community-based question answering as shown in \cite{zhao2016expert} and their ranking metric network learning framework and in \cite{zhao2015cold} which adresses the cold-start expert finding problem.

\section{Framework for Expert Finding Evaluation}
\label{sec:definition}

In this Section, after formally describing the expert finding task, we present two methodologies to generate queries. The first, usually used in the literature, directly sets topics labels as queries whereas the second, which we introduce in this paper, samples documents from the experts of each topic. Finally we detail how we used the data from AMiner to generate two datasets for the expert finding task. 

\subsection{Formal Description}

Let $G=(V,E)$ be a bipartite graph with nodes $V = V_C \cup V_D$ corresponding to a set of candidates $C$ and a set of documents $D$, where the links are undirected associations candidate-documents. Let $X \in \mathbb{R}^{|D| \times N}$ be the textual features of the $D$ documents. The expert finding task, given such $(G,X)$ dataset (see Figure \ref{fig:data}), consists in scoring the set of candidates given a textual query $q \in \mathbb{R}^{N}$, in order to answer the question \textit{\quotes{who are the candidates more likely to be experts in the topics present in the query ?}}. Given a set of queries $Q = (q_1, ..., q_i, ..., q_M)$ each associated with an identified set of experts $E_i \subset E \subset C$, $E$ being the global set of known experts among the candidates, we want to optimize the ranking of the ground truth experts $E_i$ among the global set of experts $E$. 

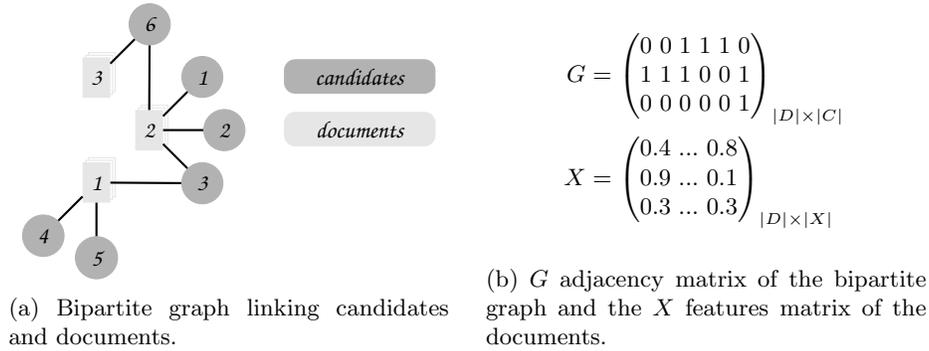
\begin{figure}[t!]
\centering
    \begin{subfigure}[b]{0.48\textwidth}
        \centering
        \begin{tikzpicture}[auto, thick, scale=0.35]
          \edef\mya{0}
          \foreach \place/\name in {{(0,-2)/a}, {(2,0)/b}, {(0,2)/d}}
           \pgfmathparse{int(\mya+1}
            \xdef\mya{\pgfmathresult}
            \node[superpeers] (\name) at \place {\mya};
           \foreach \pos/\i in {above right of/1, right of/2, below right of/3}
            \node[peers, \pos =b ] (b\i) {\i};
           \foreach \speer/\peer in {b/b1,b/b2,b/b3}
           \path (\speer) edge[-] (\peer);
           \path (a) edge[-] (b3);
           \node[peers, above right of=d] (d1){6};
           \path (d) edge[-] (d1);
           \path (b) edge[-] (d1);
           \edef\mya{3}
           \foreach \pos/\i in {below left of/1, below of/2}
            \pgfmathparse{int(\i+3)}
            \edef\mya{\pgfmathresult}
            \node[peers, \pos =a ] (a\i) {\mya};
           \foreach \speer/\peer in {a/a1,a/a2}
           \path (\speer) edge[-] (\peer);
           \node[legendsp] at (10,0) {\small{documents}};
           \node[legendp] at (10,2) {\small{candidates}};
        \end{tikzpicture}
        \caption{Bipartite graph linking candidates and documents.}
        \label{fig:graph}
\end{subfigure}%
~~~~
\begin{subfigure}[b]{0.48\textwidth}
        \centering
        \begin{align*}
            G &= \begin{pmatrix}
              0        & 0       & 1       & 1         & 1        & 0\\
              1        & 1       & 1       & 0         & 0        & 1\\
              0        & 0       & 0       & 0         & 0        & 1\\
            \end{pmatrix}_{|D|\times|C|} \\
            X &= \begin{pmatrix}
              0.4        & ...   & 0.8  \\
              0.9        & ...   & 0.1  \\
              0.3        & ...   & 0.3  \\
            \end{pmatrix}_{|D|\times|X|}
        \end{align*}
        \caption{$G$ adjacency matrix of the bipartite graph and the $X$ features matrix of the documents.}
   \label{fig:data}
\end{subfigure}
\caption{Hypothetical example of a dataset for expert finding.}
\label{fig:data}
\end{figure}  

\subsection{Evaluation}

To evaluate the ranking of experts produced by an algorithm given a query, we use several common metrics from information retrieval such as Precision at rank K (P@K), Average Precision (AP) and Reciprocal Rank (RR). Moreover, to better understand the behavior of the algorithms tested, we construct the Receiver Operating Characteristic (ROC) curve and compute its Area Under the Curve (AUC). For each of these metrics, we also compute their standard deviations along the queries which shows the robustness of the tested algorithms against the variations in the data. Moreover, when we have multiple queries per topic, we compute the standard deviation along the topics. We now present two ways of generating queries and their corresponding ground truth experts.      

\subsubsection{\textit{Topic-query} evaluation}

This approach is straightforward and is commonly adopted in the expert finding community. For a specific topic, its naming or description is directly used as a query and its associated experts are the ground truth list of candidates to be retrieved. Algorithm \ref{code:topics} shows the complete evaluation procedure. As a result, if the dataset is composed of 10 topics, the protocol of evaluation consists of 10 queries. We call this approach the \textit{topic-query} evaluation. For each measure described above, we are interested in its mean (\textit{Mean}) and standard deviation (\textit{STD}) along the queries.    

\begin{algorithm}
\caption{\textit{Topic-query} evaluation procedure. The function \textit{Evaluate} generates metrics such as P@10 and the ROC AUC based on the produced ranking and the ground truth expert set of a given topic.}
\label{code:topics}
\begin{algorithmic}
\REQUIRE Ranking\_Algorithm
\STATE scores $\leftarrow$ [~]
\FORALL{topics} 
\STATE candidates\_ranking = Ranking\_Algorithm(current\_topic\_textual\_expression)
\STATE current\_score $\leftarrow$ Evaluate(candidates\_ranking, ground\_truth\_experts\_set)
\STATE scores.append(current\_score) 
\ENDFOR
\RETURN Mean(scores), STD(scores)
\end{algorithmic}
\end{algorithm}
 
\subsubsection{\textit{Document-query} evaluation} 

We propose to sample the documents linked with the experts of a given topic in order to use them as queries. Instead of using the topic description, we use the set of documents associated to the ground truth experts of a given topic. Precisely, we create a set of queries and their associated experts by selecting each document of the dataset linked with the ground truth experts. As such, the evaluated algorithm produces a ranked list of candidates for each document-query and its performance is measured by comparing the ranking with the experts of the same topic as the expert who produced the document-query. Since several document-queries are sampled for each topics, we also compute the means and standard deviations along the topics, by computing these values along the averaged measures intra-topics. To avoid any bias in the metrics, when evaluating an algorithm on a sampled document, we leave it out of the data. We call this approach \textit{document-query} evaluation. Algorithm \ref{code:documents} shows the complete evaluation procedure.

\begin{algorithm}
\caption{\textit{Document-query} evaluation procedure. Note that the computed metrics are also averaged for each topic in order to compute the inter-topic standard deviation.}
\label{code:documents}
\begin{algorithmic}
\REQUIRE Ranking\_Algorithm
\STATE scores $\leftarrow$ [~]
\STATE topical\_scores $\leftarrow$ \{\}
\FORALL{topics}
\STATE topical\_scores[current\_topic] $\leftarrow$ [~]
\FORALL{experts\_of\_current\_topic}
\FORALL{documents\_of\_current\_expert}
\STATE candidates\_ranking = Ranking\_Algorithm(current\_document\_textual\_expression,
\STATE ~~~~~~~~~~~~~~~~~~~~~~~~~~~~~~~~~~~~~~~~~~~~~~~~~~~~~~~leave\_out = current\_document)
\STATE current\_score $\leftarrow$ Evaluate(candidates\_ranking, ground\_truth\_experts\_set)
\STATE scores.append(current\_score) 
\STATE topical\_scores[topics].append(current\_score)
\ENDFOR
\ENDFOR
\STATE topical\_scores[current\_topic] $\leftarrow$ Mean(topical\_scores[current\_topic])
\ENDFOR
\RETURN Mean(scores), STD(scores), STD(topical\_scores)
\end{algorithmic}
\end{algorithm}      

\subsection{AMiner Data}
The AMiner project aims to provide tools for mining researcher's social network. They provided several datasets \footnote{https://AMiner.org/data} \cite{tang2008arnetminer}~collecting papers, authors, co-authorship and citations links extracted from DBLP  \cite{ley2002dblp}, ACM (Association for Computing Machinery) and other sources in the field of computer science. For the task of expert finding, they provided two lists of experts \footnote{https://AMiner.org/lab-datasets/expertfinding/\#expert-list}. The first, the \textit{machine-annotated list}, is composed of 13 topics and has been built from topical web search. The second, the \textit{human-annotated list}, is composed of 7 topics built with the method of pooled relevance judgments together with human judgments as described in \cite{zhang2008mixture}. We used the \textit{machine-annotated list} with the \textit{citation dataset V2} and the \textit{human-annotated list} with the \textit{citation dataset V1} available on the AMiner website \footnote{https://AMiner.org/citation}.  

We preprocessed the two datasets based on the distribution of links between candidates and documents. We also took into account the document string length (number of letters). First we kept only authors with less than 100 documents links and with at least one link. This reduces author name ambiguity by discarding authors who were originally connected to tens of thousands documents. Then we composed the textual content of the documents by concatenating their titles and abstracts and by keeping only those with string length greater than 50. As a result, we ended up with two datasets:
\begin{itemize}
\item AMiner expert dataset 1: using the \textit{machine-annotated list} of experts, is composed of 996,110 candidates, 1,125,082 documents, 1,269 experts in 13 topics. The distribution of the experts across topics is given in Table \ref{table:experts1} (one expert can be linked to several topics) with the total number of documents linked to those experts
\item AMiner expert dataset 2: using the \textit{human-annotated list} of experts, is composed of 532,968 candidates, 480,630 documents, 210 experts in 7 topics. The distribution is given in Table \ref{table:experts2}.
\end{itemize}

\begin{table}[t!]
\centering
    \begin{subtable}[t]{0.48\textwidth}
            \footnotesize
            \caption{AMiner dataset 1.}
            \label{table:experts1}
            \begin{tabular}{|c|c|c|}
            \hline
            \textbf{Topic} & \textbf{Exps} & \textbf{Docs} \\ \hline
            neural networks          &105           &1941           \\ 
            ontology alignment          &49           &908           \\ 
            boosting          &49           &1062           \\ 
            support vector machine          &91           &1145           \\ 
            intelligent agents          &28           &729           \\ 
            machine learning          &33           &1229           \\ 
            computer vision          &174           &3636           \\ 
            data mining          &285           &8034           \\ 
            natural language processing         &40           &1480           \\ 
            semantic web          &332           &6435           \\ 
            planning          &21           &531           \\ 
            cryptography          &139           &3033           \\ 
            \textbf{TOTAL} & \textbf{1269} & \textbf{30802}  \\ \hline
            \end{tabular}
    \end{subtable}
    \hspace{\fill} 
    \begin{subtable}[t]{0.48\textwidth}
        \flushright
        \footnotesize
            \caption{AMiner dataset 2.}
            \label{table:experts2}
            \begin{tabular}{|c|c|c|}
            \hline
            \textbf{Topic} & \textbf{Exps} & \textbf{Docs} \\ \hline
            intelligent agents           &29           &685           \\ 
            planning          &33           &510           \\ 
            machine learning          &38           &628           \\ 
            natural language processing          &38           &453           \\ 
            information extraction          &18           &290           \\ 
            semantic web          &41           &521           \\ 
            support vector machine          &29           &323           \\ 
            \textbf{TOTAL} & \textbf{210} & \textbf{3410}  \\ \hline
            \end{tabular}
    \end{subtable}
\caption{Distribution of experts and their documents counts across topics in both AMiner expert datasets.}
\label{fig:experts}
\end{table} 

\section{Baseline Algorithms} 

\label{sec:baseline}

After a short description of document representation, we describe three baseline algorithms taken from the literature. We reimplemented them since their original codes were not available or hardly reusable. Moreover, we could easily extend them to work with any kind of document representation.

\subsection{Document Representation}

Our three baseline algorithms rely on a measure of semantic similarity between the queries and the corpus of documents. We chose to try several document representations: term frequency (TF), term frequency - inverse document frequency (TF-IDF) and latent semantic indexing (LSI) \cite{papadimitriou1998latent}. We tokenized the text of the documents by lowercasing the characters, removing stop words and concatenating tokens based on their co-occurrence counts to compound 2-grams and 3-grams. Then, words appearing less than 3 times in the corpus or in more than 50\% of the documents were discarded to reduce the computational cost without affecting the retrieval performance. The number of dimensions of the singular value decomposition for the LSI is 300. This number was chosen to ensure components above noise level are retained as proposed in \cite{kaiser1960application}. 

\subsection{Text-based Approach 1: P@noptic Model}

\textit{P@noptic Expert} \cite{craswell2001p} is a simple algorithm which creates meta-documents for each author. Our implementation first concatenates the contents (title+abstract) of all documents linked with each candidate, then vectorizes this meta-documents using the pretrained documents representation models. Finally, it computes the cosine similarities between a query and the meta-documents and ranks the candidates by descending order of their scores.   

\subsection{Text-based Approach 2: Voting Model}

Our voting model based on \cite{macdonald2006voting} first computes the cosine similarities between the query and the documents of the dataset and then ranks all documents by descending order of their score. The algorithm then sums the inverse value of the rank (Reciprocal Rank - RR) of each document a candidate is linked with. If a candidate is linked with the 2nd, 3rd and 7th closest documents to the query, its score will be $\frac{1}{2} + \frac{1}{3} + \frac{1}{7} = 0.976$. This algorithm gives a huge boost to candidates who have at least one document well ranked and tends to promote candidates with more documents than others. We also tried other fusion techniques than the RR such as CombSUM and CombMNZ, described in \cite{macdonald2006voting}, but they provided weaker results. 

\subsection{Graph and Text-based Approach: Propagation Model}

The propagation model we made is a simpler version of those described in \cite{serdyukov2008modeling}. The algorithm first computes the cosine similarities between the query and the documents and it initializes a score vector $S_0$ of length $|C|+|D|$ with zeros for candidates and the documents-query scores for documents. It then operates several two-steps random walks with restart until the score vector converges (until the $L_2$ norm of the difference of its previous value and current value is below $10^{-6}$). These random walks are done iteratively: $S_{i+1} = (1-\eta) \times A (A S_i) + \eta R$ where $\eta$ is the jumping factor, a scalar between 0 and 1, which controls the restart, $R = S_0$ is the restart vector that represents the global probability of a random walk to restart from its original node, $A$ is the column-wise $L_1$ normalized adjacency matrix of the bipartite graph, also known as the PageRank transition matrix \cite{page1999pagerank}. At each step, scores jump from documents to candidates then from candidates to documents. A last step is finally done to propagate scores back to the candidates. These scores are then ranked by descending order.

\section{Experiments}

\label{sec:experiments}

In this section, we present the experiments we did with both \textit{topic-query} and \textit{document-query} evaluations. We first show some general results before analyzing the effect of the type of query and finally focusing on the variations of ranking along the queries and the topics.

\subsection{Settings}

We evaluated our baseline models on the \textit{topic-query} and the \textit{document-query} methodologies. We made two evaluations for the propagation model using $\eta=0.1$ where the restart is weak, hence the propagation is wide, and $\eta=0.5$ where the scores stay close to their initial values. Moreover, for each model, the semantic similarity was computed with TF, TF-IDF and LSI document representations. Table \ref{tab:results1} shows the results on the AMiner expert dataset 1 and Table \ref{tab:results2} shows the results on the AMiner expert dataset 2.

\subsection{General Results}

For both datasets, the document representation TF-IDF performs generally better except for the AUC score, where LSI performs best, especially on \textit{topic queries}.
Actually, taking a closer look at the ROC curve, we could see that LSI is better in ranking for the worst ranked experts.
It smoothes the curve in the top right corner and hence improves the area under the curve. Most metrics (P@10 and RR for example) are intended to focus on the quality of the very first ranked experts but the ROC AUC allows us to analyze the behavior of a ranking algorithm over the entire ranking. It is also important to note that the results are more stable across the choice of document representation for the second dataset.
This behavior is expected since the ground truth experts have been human curated.

\subsection{Effect of the Type of Query}

We observe different rankings of the baseline algorithms depending on the type of evaluation performed. For the \textit{topic-query} procedure, the propagation model performs best (with $\eta=0.5$ for the first dataset and $\eta=0.1$ for the second) whereas the voting model is the best for the \textit{document-query} evaluation.
Our explanation is that voting models are good when queries and documents are of the same type since we only need one candidate's document to be similar to the query to push her to the top of the ranking. When the query is as short as \quotes{data mining}, the chance to find such a similar document is low since few documents about data mining have the words \quotes{data mining} in their content. Indeed, scientific articles rarely deal with data mining in general but rather focus on particular aspect of this field.  

In contrast, the propagation model can give a good score to a candidate if in her neighborhood, the query is similar to some documents. Even if this candidate is an expert of \quotes{data mining} without never actually using the expression, there are quite some chances that in its close social network, some other candidates used these two words.

Then, the voting model might perform best than propagation for document queries because the latter tends to mistaken an information retrieval expert who worked closely with data mining experts. This situation is less likely to happen when the query is a short and very specific description than with paper contents that share a lot of similar terms between topics.

Moreover, this difference of results between query types are weaker when using LSI, which is due to the ability of this document representation to capture a similarity between two texts that do not share any word in common. The effect of short query is thus highly reduced compared to TF and TF-IDF.

\subsection{Standard Deviation along Queries and Topics}

One important aspect is the amount of dispersion the sets of scores have around their means. We computed the standard deviations for each evaluation to have an insight of the robustness of the algorithms to queries and to topics. Interestingly, for the \textit{document-query}, the standard deviations along topics evaluation are lower than the deviations along queries. This shows that the robustness of the algorithms are not that much impacted by the variation of topics, as could have suggested the standard deviations for the \textit{topic-query} evaluation, but merely by the variety of queries intra-topics. As a result, using only a few topic queries is statistically biased since some topic namings might have lesser chance to appear in their related documents. Finally, in the second dataset, the deviations along topics for the voting model are significantly lower than other models which is a precious information that cannot be revealed by a \textit{topic-query} evaluation if one wants to favorite stability over the searched topics of expertise.

\subsection{Pros and Cons of the \textit{Document-Query} Evaluation} 

Beside the fact that the \textit{document-query} evaluation seems to better represent a real case application of expert finding, we showed that it provides a deeper insight on the robustness of an algorithm. The different rankings of algorithms for both evaluations and their corresponding inter-topics standard deviations prove that using only the namings of the topics is not a satisfactory protocol to compare expert finding systems. However, in a general manner, measures are much better with the \textit{topic-query} evaluation. This is due to two aspects:
\begin{itemize}
\item document-queries are semantically fine grained and it is more difficult to separate two queries of different topics. This makes the expert finding task harder to solve but it is not a bad thing for the evaluation.   
\item in our current configuration, document-queries do not rely on an annotated dataset. As a consequence, some sampled documents might not actually belong to the topic their authors are associated with. This motivates the construction of a ground truth set of documents associated to at least one of the human-annotated expert topics.  
\end{itemize}

\afterpage{\clearpage}
\begin{table}[p]
\centering
\footnotesize
    \begin{subtable}[t]{1\textwidth}
    \centering
        \caption{Baseline mean scores and their query (same as topic) standard deviations for the \textit{topic-query} evaluation.}
        \label{tab:inter-topic-1}
        \begin{tabular}{cc|c|c|c|}
        \cline{3-5}  &  &TF  &TF-IDF  &LSI  \\ \hline
        \multicolumn{1}{|l|}{\multirow{4}{*}{P@noptic}} & AUC  & 0.777$\pm$0.099 & 0.778$\pm$0.102 & 0.832$\pm$0.083\\
        \multicolumn{1}{|l|}{} & P@10 & 0.662$\pm$0.262 & \textbf{0.685}$\pm$0.260 & \textbf{0.615}$\pm$0.301\\
        \multicolumn{1}{|l|}{} & AP & 0.398$\pm$0.193 & 0.415$\pm$0.204 & \textbf{0.395}$\pm$0.233\\
        \multicolumn{1}{|l|}{} & RR & 1.615$\pm$0.923 & 1.538$\pm$0.746 & 1.769$\pm$0.890\\ \hline
        \multicolumn{1}{|l|}{\multirow{4}{*}{Vote}} & AUC  & 0.714$\pm$0.137 & 0.714$\pm$0.138 & 0.800$\pm$0.107\\
        \multicolumn{1}{|l|}{} & P@10 & 0.608$\pm$0.312 & 0.608$\pm$0.287 & 0.538$\pm$0.325\\
        \multicolumn{1}{|l|}{} & AP & 0.373$\pm$0.212 & 0.381$\pm$0.209 & 0.390$\pm$0.247\\
        \multicolumn{1}{|l|}{} & RR & 2.308$\pm$2.398 & 1.538$\pm$0.929 & 2.769$\pm$3.765\\ \hline
        \multicolumn{1}{|l|}{\multirow{4}{*}{\shortstack{Prop \\ ($\eta = 0.1$)}}} & AUC  & 0.834$\pm$0.093 & 0.834$\pm$0.096 & 0.824$\pm$0.085\\
        \multicolumn{1}{|l|}{} & P@10 & 0.669$\pm$0.270 & 0.677$\pm$0.264 & \textbf{0.615}$\pm$0.298\\
        \multicolumn{1}{|l|}{} & AP & \textbf{0.458}$\pm$0.239 & \textbf{0.473}$\pm$0.243 & 0.389$\pm$0.230\\
        \multicolumn{1}{|l|}{} & RR & 1.462$\pm$0.929 & 1.538$\pm$1.082 & \textbf{1.692}$\pm$1.136\\ \hline
        \multicolumn{1}{|l|}{\multirow{4}{*}{\shortstack{Propagation \\ ($\eta = 0.5$)}}} & AUC  & \textbf{0.842}$\pm$0.086 & \textbf{0.842}$\pm$0.088 & \textbf{0.833}$\pm$0.083\\
        \multicolumn{1}{|l|}{} & P@10 & \textbf{0.677}$\pm$0.255 & \textbf{0.685}$\pm$0.274 & \textbf{0.615}$\pm$0.296\\
        \multicolumn{1}{|l|}{} & AP & 0.457$\pm$0.232 & 0.472$\pm$0.242 & \textbf{0.395}$\pm$0.231\\
        \multicolumn{1}{|l|}{} & RR & \textbf{1.308}$\pm$0.606 & \textbf{1.385}$\pm$0.738 & \textbf{1.692}$\pm$1.136\\ \hline

        \end{tabular}
    \end{subtable}
        \small
    \begin{subtable}[t]{0.68\textwidth}
    \centering
        \caption{Baseline mean scores and their query standard deviations for the \textit{document-query} evaluation.}
        \label{tab:inter-documents-1}
        \begin{tabular}{cc|c|c|c|}
        \cline{3-5}  &  &TF  &TF-IDF  &LSI  \\ \hline
        \multicolumn{1}{|l|}{\multirow{4}{*}{P@noptic}} & AUC  & 0.593$\pm$0.131 & 0.618$\pm$0.134 & 0.620$\pm$0.139\\
        \multicolumn{1}{|l|}{} & P@10 & 0.255$\pm$0.266 & 0.335$\pm$0.294 & \textbf{0.302}$\pm$0.298\\
        \multicolumn{1}{|l|}{} & AP & 0.150$\pm$0.117 & 0.181$\pm$0.133 & 0.174$\pm$0.133\\
        \multicolumn{1}{|l|}{} & RR & 9.153$\pm$16.028 & 6.210$\pm$13.323 & 8.663$\pm$16.262\\ \hline
        \multicolumn{1}{|l|}{\multirow{4}{*}{Vote}} & AUC  & \textbf{0.606}$\pm$0.131 & \textbf{0.630}$\pm$0.137 & \textbf{0.637}$\pm$0.142\\
        \multicolumn{1}{|l|}{} & P@10 & \textbf{0.284}$\pm$0.263 & 0.322$\pm$0.280 & 0.275$\pm$0.276\\
        \multicolumn{1}{|l|}{} & AP & \textbf{0.169}$\pm$0.131 & \textbf{0.193}$\pm$0.145 & \textbf{0.187}$\pm$0.152\\
        \multicolumn{1}{|l|}{} & RR & \textbf{6.819}$\pm$15.421 & \textbf{5.783}$\pm$13.983 & \textbf{8.438}$\pm$16.987\\ \hline
        \multicolumn{1}{|l|}{\multirow{4}{*}{\shortstack{Propagation \\ ($\eta = 0.1$)}}} & AUC  & 0.591$\pm$0.140 & 0.617$\pm$0.143 & 0.612$\pm$0.145\\
        \multicolumn{1}{|l|}{} & P@10 & 0.256$\pm$0.253 & \textbf{0.336}$\pm$0.292 & 0.300$\pm$0.288\\
        \multicolumn{1}{|l|}{} & AP & 0.152$\pm$0.114 & 0.183$\pm$0.132 & 0.173$\pm$0.127\\
        \multicolumn{1}{|l|}{} & RR & 9.035$\pm$16.718 & 6.773$\pm$14.641 & 8.948$\pm$17.422\\ \hline
        \multicolumn{1}{|l|}{\multirow{4}{*}{\shortstack{Propagation \\ ($\eta = 0.5$)}}} & AUC  & 0.598$\pm$0.141 & 0.623$\pm$0.142 & 0.621$\pm$0.147\\
        \multicolumn{1}{|l|}{} & P@10 & 0.255$\pm$0.245 & 0.333$\pm$0.283 & 0.298$\pm$0.282\\
        \multicolumn{1}{|l|}{} & AP & 0.155$\pm$0.115 & 0.185$\pm$0.133 & 0.177$\pm$0.130\\
        \multicolumn{1}{|l|}{} & RR & 8.994$\pm$17.224 & 6.419$\pm$14.088 & 9.089$\pm$18.358\\ \hline
        \end{tabular}
    \end{subtable}
    \hspace{\fill} 
    \begin{subtable}[t]{0.27\textwidth}
    \centering
    \flushright
        \caption{Topic standard deviations.}
        \label{tab:intra-documents-1}
        \begin{tabular}{|c|c|c|}
        \hline
         TF  &TF-IDF  &LSI  \\ \hline
        0.114 & 0.112 & 0.119\\
        0.174 & 0.191 & 0.195\\
        0.097 & 0.102 & 0.104\\
        6.063 & 3.676 & 6.088\\ \hline
        \textit{0.109} & \textit{0.109} & \textit{0.111}\\
        0.166 & 0.181 & \textit{0.174}\\
        0.102 & 0.110 & 0.115\\
        \textit{3.400} & \textit{2.789} & \textit{4.168}\\ \hline
        0.128 & 0.126 & 0.131\\
        0.156 & 0.187 & 0.184\\
        \textit{0.093} & 0.102 & \textit{0.101}\\
        9.175 & 6.139 & 9.245\\ \hline
        0.125 & 0.122 & 0.128\\
        \textit{0.147} & \textit{0.177} & 0.175\\
        \textit{0.093} & \textit{0.101} & 0.102\\
        8.488 & 5.059 & 8.944\\ \hline
        \end{tabular}
    \end{subtable}
\caption{Results of the evaluations with the AMiner dataset 1, composed with the machine-annotated experts and the biggest candidates-documents set. Bold values are the best scores across the algorithms for each document representation.}
\label{tab:results1}
\end{table} 

\afterpage{\clearpage}
\begin{table}[p]
\centering
    \begin{subtable}[t]{1\textwidth}
    \centering
        \caption{Baseline mean scores and their query (same as topic) standard deviations for the \textit{topic-query} evaluation.}
        \label{tab:inter-topic-1}
        \begin{tabular}{cc|c|c|c|}
        \cline{3-5}  &  &TF  &TF-IDF  &LSI  \\ \hline
        \multicolumn{1}{|l|}{\multirow{4}{*}{P@noptic}} & AUC  & 0.809$\pm$0.114 & 0.815$\pm$0.119 & 0.853$\pm$0.059\\
        \multicolumn{1}{|l|}{} & P@10 & 0.757$\pm$0.176 & 0.814$\pm$0.146 & \textbf{0.771}$\pm$0.183\\
        \multicolumn{1}{|l|}{} & AP & 0.580$\pm$0.175 & 0.613$\pm$0.186 & 0.580$\pm$0.157\\
        \multicolumn{1}{|l|}{} & RR & \textbf{1.000}$\pm$0.000 & \textbf{1.000}$\pm$0.000 & 1.429$\pm$0.495\\ \hline
        \multicolumn{1}{|l|}{\multirow{4}{*}{Vote}} & AUC  & 0.788$\pm$0.131 & 0.793$\pm$0.136 & \textbf{0.857}$\pm$0.048\\
        \multicolumn{1}{|l|}{} & P@10 & 0.786$\pm$0.136 & 0.800$\pm$0.141 & 0.729$\pm$0.158\\
        \multicolumn{1}{|l|}{} & AP & 0.607$\pm$0.158 & 0.636$\pm$0.171 & \textbf{0.599}$\pm$0.123\\
        \multicolumn{1}{|l|}{} & RR & 1.286$\pm$0.452 & \textbf{1.000}$\pm$0.000 & \textbf{1.000}$\pm$0.000\\ \hline
        \multicolumn{1}{|l|}{\multirow{4}{*}{\shortstack{Prop \\ ($\eta = 0.1$)}}} & AUC  & \textbf{0.860}$\pm$0.097 & \textbf{0.866}$\pm$0.100 & 0.834$\pm$0.052\\
        \multicolumn{1}{|l|}{} & P@10 & \textbf{0.829}$\pm$0.148 & \textbf{0.843}$\pm$0.118 & 0.686$\pm$0.181\\
        \multicolumn{1}{|l|}{} & AP & \textbf{0.647}$\pm$0.142 & \textbf{0.676}$\pm$0.139 & 0.564$\pm$0.140\\
        \multicolumn{1}{|l|}{} & RR & \textbf{1.000}$\pm$0.000 & \textbf{1.000}$\pm$0.000 & 1.143$\pm$0.350\\ \hline
        \multicolumn{1}{|l|}{\multirow{4}{*}{\shortstack{Prop \\ ($\eta = 0.5$)}}} & AUC  & \textbf{0.860}$\pm$0.098 & 0.864$\pm$0.101 & 0.843$\pm$0.057\\
        \multicolumn{1}{|l|}{} & P@10 & 0.786$\pm$0.146 & 0.814$\pm$0.125 & 0.686$\pm$0.181\\
        \multicolumn{1}{|l|}{} & AP & 0.646$\pm$0.139 & 0.671$\pm$0.140 & 0.579$\pm$0.138\\
        \multicolumn{1}{|l|}{} & RR & \textbf{1.000}$\pm$0.000 & 1.000$\pm$0.000 & 1.286$\pm$0.452\\ \hline

        \end{tabular}
    \end{subtable}
    \begin{subtable}[t]{0.68\textwidth}
    \centering
        \caption{Baseline mean scores and their query standard deviations for the \textit{document-query} evaluation.}
        \label{tab:inter-documents-1}
        \begin{tabular}{cc|c|c|c|}
        \cline{3-5}  &  &TF  &TF-IDF  &LSI  \\ \hline
       \multicolumn{1}{|l|}{\multirow{4}{*}{P@noptic}} & AUC  & 0.599$\pm$0.112 & 0.626$\pm$0.123 & 0.621$\pm$0.121\\
        \multicolumn{1}{|l|}{} & P@10 & 0.324$\pm$0.278 & 0.387$\pm$0.285 & 0.343$\pm$0.287\\
        \multicolumn{1}{|l|}{} & AP & 0.282$\pm$0.145 & 0.318$\pm$0.164 & 0.302$\pm$0.158\\
        \multicolumn{1}{|l|}{} & RR & 4.904$\pm$5.731 & 3.557$\pm$4.976 & 4.810$\pm$5.895\\ \hline
        \multicolumn{1}{|l|}{\multirow{4}{*}{Vote}} & AUC  & \textbf{0.611}$\pm$0.120 & \textbf{0.637}$\pm$0.129 & \textbf{0.634}$\pm$0.132\\
        \multicolumn{1}{|l|}{} & P@10 & \textbf{0.370}$\pm$0.257 & \textbf{0.417}$\pm$0.274 & \textbf{0.370}$\pm$0.266\\
        \multicolumn{1}{|l|}{} & AP & \textbf{0.303}$\pm$0.148 & \textbf{0.338}$\pm$0.168 & \textbf{0.318}$\pm$0.161\\
        \multicolumn{1}{|l|}{} & RR & \textbf{3.211}$\pm$4.637 & \textbf{2.752}$\pm$4.211 & \textbf{3.686}$\pm$5.174\\ \hline
        \multicolumn{1}{|l|}{\multirow{4}{*}{\shortstack{Propagation \\ ($\eta = 0.1$)}}} & AUC  & 0.596$\pm$0.113 & 0.625$\pm$0.121 & 0.612$\pm$0.119\\
        \multicolumn{1}{|l|}{} & P@10 & 0.325$\pm$0.269 & 0.381$\pm$0.283 & 0.339$\pm$0.278\\
        \multicolumn{1}{|l|}{} & AP & 0.283$\pm$0.147 & 0.319$\pm$0.163 & 0.298$\pm$0.158\\
        \multicolumn{1}{|l|}{} & RR & 4.512$\pm$5.510 & 3.557$\pm$5.171 & 4.558$\pm$6.141\\ \hline
        \multicolumn{1}{|l|}{\multirow{4}{*}{\shortstack{Propagation \\ ($\eta = 0.5$)}}} & AUC  & 0.598$\pm$0.114 & 0.624$\pm$0.122 & 0.615$\pm$0.121\\
        \multicolumn{1}{|l|}{} & P@10 & 0.335$\pm$0.268 & 0.392$\pm$0.280 & 0.351$\pm$0.279\\
        \multicolumn{1}{|l|}{} & AP & 0.286$\pm$0.146 & 0.320$\pm$0.162 & 0.303$\pm$0.158\\
        \multicolumn{1}{|l|}{} & RR & 4.027$\pm$5.039 & 3.280$\pm$4.885 & 4.197$\pm$5.902\\ \hline

        \end{tabular}
    \end{subtable}
    \hspace{\fill} 
    \begin{subtable}[t]{0.27\textwidth}
    \centering
    \flushright
        \caption{Topic standard deviations.}
        \label{tab:intra-documents-1}
        \begin{tabular}{|c|c|c|}
        \hline
         TF  &TF-IDF  &LSI  \\ \hline
        0.061 & 0.058 & 0.064\\
        0.194 & 0.171 & 0.195\\
        0.095 & 0.094 & 0.101\\
        3.265 & 2.019 & 3.142\\ \hline
        \textit{0.059} & \textit{0.056} & \textit{0.058}\\
        \textit{0.110} & \textit{0.112} & \textit{0.115}\\
        \textit{0.067} & \textit{0.073} & \textit{0.071}\\
        \textit{1.107} & \textit{0.908} & \textit{1.307}\\ \hline
        0.077 & 0.074 & 0.079\\
        0.190 & 0.181 & 0.195\\
        0.101 & 0.104 & 0.108\\
        3.120 & 2.206 & 3.320\\ \hline
        0.073 & 0.068 & 0.076\\
        0.178 & 0.164 & 0.181\\
        0.094 & 0.094 & 0.101\\
        2.414 & 1.730 & 2.856\\ \hline
        \end{tabular}
    \end{subtable}
\caption{Results of the evaluations with the AMiner dataset 2, composed with the human-annotated experts and the smallest candidates-documents set. Bold values are the best scores across the algorithms for each document representation.}
\label{tab:results2}
\end{table}

\section{Summary and Future Work}

We compared two evaluation protocols for scientific expert finding that rely on two types of query generation. Evaluating our baseline models with this framework, we showed that using the documents written by the ground truth experts brings different results than with the usual topic queries. Specifically, short queries can profit of a propagation model whereas longer queries are better handled by a simpler voting model. Moreover, the lower standard deviations along topics for the \textit{document-query} evaluation shows that there is a bias in using only one topic naming as query since the document representations do not handle well such short query similarity to the documents.

To improve the \textit{document-query} evaluation with the AMiner data, we would like to filter the set of sampled documents by human annotation in order to keep only those that match the expertise of their authors. This would then justify a deeper analysis of the significance of the measurements to consider the variations of ranking of the evaluated algorithms along the queries. Another interesting work would be to perform an online evaluation of the same expert finding algorithms in the case of a reviewer assignment application in order to compare the results with our framework.

\clearpage
%
%
%
\bibliographystyle{splncs04}
\bibliography{biblio}

\begin{thebibliography}{10}
\providecommand{\url}[1]{\texttt{#1}}
\providecommand{\urlprefix}{URL }
\providecommand{\doi}[1]{https://doi.org/#1}

\bibitem{balog2006formal}
Balog, K., Azzopardi, L., De~Rijke, M.: Formal models for expert finding in
  enterprise corpora. In: Proceedings of the 29th annual international ACM
  SIGIR conference on Research and development in information retrieval. pp.
  43--50. ACM (2006)

\bibitem{craswell2001p}
Craswell, N., Hawking, D., Vercoustre, A.M., Wilkins, P.: P@ noptic expert:
  Searching for experts not just for documents. In: Ausweb Poster Proceedings,
  Queensland, Australia. vol.~15, p.~17 (2001)

\bibitem{craswell2005overview}
Craswell, N., de~Vries, A.P., Soboroff, I.: Overview of the trec 2005
  enterprise track. In: Trec. vol.~5, pp. 199--205 (2005)

\bibitem{dong2017metapath2vec}
Dong, Y., Chawla, N.V., Swami, A.: metapath2vec: Scalable representation
  learning for heterogeneous networks. In: Proceedings of the 23rd ACM SIGKDD
  International Conference on Knowledge Discovery and Data Mining. pp.
  135--144. ACM (2017)

\bibitem{kaiser1960application}
Kaiser, H.F.: The application of electronic computers to factor analysis.
  Educational and psychological measurement  \textbf{20}(1),  141--151 (1960)

\bibitem{ley2002dblp}
Ley, M.: The dblp computer science bibliography: Evolution, research issues,
  perspectives. In: International symposium on string processing and
  information retrieval. pp. 1--10. Springer (2002)

\bibitem{macdonald2006voting}
Macdonald, C., Ounis, I.: Voting for candidates: adapting data fusion
  techniques for an expert search task. In: Proceedings of the 15th ACM
  international conference on Information and knowledge management. pp.
  387--396. ACM (2006)

\bibitem{page1999pagerank}
Page, L., Brin, S., Motwani, R., Winograd, T.: The pagerank citation ranking:
  Bringing order to the web. Tech. rep., Stanford InfoLab (1999)

\bibitem{papadimitriou1998latent}
Papadimitriou, C.H., Tamaki, H., Raghavan, P., Vempala, S.: Latent semantic
  indexing: A probabilistic analysis. In: Proceedings of the seventeenth ACM
  SIGACT-SIGMOD-SIGART symposium on Principles of database systems. pp.
  159--168. ACM (1998)

\bibitem{serdyukov2008modeling}
Serdyukov, P., Rode, H., Hiemstra, D.: Modeling multi-step relevance
  propagation for expert finding. In: Proceedings of the 17th ACM conference on
  Information and knowledge management. pp. 1133--1142. ACM (2008)

\bibitem{tang2008arnetminer}
Tang, J., Zhang, J., Yao, L., Li, J., Zhang, L., Su, Z.: Arnetminer: extraction
  and mining of academic social networks. In: Proceedings of the 14th ACM
  SIGKDD international conference on Knowledge discovery and data mining. pp.
  990--998. ACM (2008)

\bibitem{van2016unsupervised}
Van~Gysel, C., de~Rijke, M., Worring, M.: Unsupervised, efficient and semantic
  expertise retrieval. In: Proceedings of the 25th International Conference on
  World Wide Web. pp. 1069--1079. International World Wide Web Conferences
  Steering Committee (2016)

\bibitem{yang2015network}
Yang, C., Liu, Z., Zhao, D., Sun, M., Chang, E.Y.: Network representation
  learning with rich text information. In: IJCAI. pp. 2111--2117 (2015)

\bibitem{zhang2007expert}
Zhang, J., Tang, J., Li, J.: Expert finding in a social network. In:
  International Conference on Database Systems for Advanced Applications. pp.
  1066--1069. Springer (2007)

\bibitem{zhang2008mixture}
Zhang, J., Tang, J., Liu, L., Li, J.: A mixture model for expert finding. In:
  Pacific-Asia Conference on Knowledge Discovery and Data Mining. pp. 466--478.
  Springer (2008)

\bibitem{zhao2015cold}
Zhao, Z., Wei, F., Zhou, M., Ng, W.: Cold-start expert finding in community
  question answering via graph regularization. In: International Conference on
  Database Systems for Advanced Applications. pp. 21--38. Springer (2015)

\bibitem{zhao2016expert}
Zhao, Z., Yang, Q., Cai, D., He, X., Zhuang, Y.: Expert finding for
  community-based question answering via ranking metric network learning. In:
  IJCAI. pp. 3000--3006 (2016)

\end{thebibliography}
%
\end{document}